\documentclass[12pt]{iopart}

\newcommand{\cs}{\left| \alpha\rangle\langle\alpha \right|}
\newcommand{\al}{\alpha}

\newcommand{\pa}{\partial}

\usepackage{amssymb}
\usepackage{iopams}

\begin{document}

\title[Electron in ideal phonon gas]{Electron %
random walk in ideal phonon gas. \\
Exact dressed electron density matrix \\ %
evolution equations}

\author{Yu E Kuzovlev}

\address{Donetsk Institute
for Physics and Technology of NASU, 83114 Donetsk, Ukraine}
\ead{kuzovlev@fti.dn.ua}

\begin{abstract}
An original approach is suggested %
to analysis of full quantum Liouville %
equation for single electron (quantum particle) interacting %
with ideal phonon gas (harmonic boson thermostat). %
It is shown that under the thermodynamic limit this equation %
can be exactly reduced to a system of %
evolution equations %
connecting density matrix of the electron and %
its simplest irreducible correlations with amplitudes of %
one, two, three and more phonon modes. Possible application %
of these new equations to explanation of %
the electron mobility 1/f-type %
fluctuations in semiconductors and other media is discussed. %
The special case of electron in static disorder is also considered. %
\end{abstract}


\vspace{2pc} \noindent{\it Keywords}\,: %
Dynamical foundations of kinetics, %
Quantum kinetic equations, %
Electron-phonon interaction, %
Electron mobility fluctuations, 1/f noise

\section{Introduction}

Here I leave for a time the problem of molecular Brownian %
motion and related mobility 1/f-type fluctuations %
in classical fluids %
\cite{pro,ppro,lpro,tmf,ig,hs1,i1,p1,p3} %
and turn to similar problems %
concerning random walks of %
electric charge carriers in conducting media and %
related electronic 1/f noise \cite{bk12,bk3,i2,last}. %
Although the second subject seems much more %
interesting from practical point of view, %
the first one gave more possibilities to %
examine how and why the conventional kinetic %
approximations of statistical mechanics %
had lost so important phenomenon as 1/f noise. %
Now it is clear that the Boltzmann's %
conjectures are wrong, and %
in more correct approximations any particle %
of both dilute gas and dense fluid  %
manifests unique random diffusivity and mobility %
(what, in essence, was predicted already in \cite{kr}). %
In the light of these achievements we may try now %
revision of kinetic models of quantum statistical %
mechanics, beginning, naturally, from the quasi-free %
electron in thermally vibrating crystal lattice. %
Or, in other words, electron in phonon bath %
(or, in generalized terminology, quantum particle %
in boson field).

Of course, real crystal lattices vibrate %
anharmonically. At that the anharmonicity %
gives rise to not only relaxation of any %
particular phonon mode but also 1/f-type %
fluctuations in rate of the relaxation %
\cite{i3}. %
These fluctuations can be transmitted to %
electron's relaxation due to phonons   %
and thus to electron's mobility. %

But the electron-phonon interaction by itself %
causes relaxation of electron's velocity and %
energy. Hence, - as experience of investigation %
of the mentioned classical systems says, - %
it by itself can cause also 1/f-type fluctuations %
in the rate of this relaxation and consequently %
in diffusivity and mobility of the electron. %
Therefore it is reasonable for the beginning %
to consider electron in harmonic lattice, %
that is in ideal phonon gas. %

In this paper our purpose is to reveal %
a minimal set of relevant statistical %
characteristics which %
completely and exactly determine all %
effects of the electron-phonon interaction, %
and derive exact evolution equations for %
such characteristics. The results will %
present a necessary mathematical base %
for further investigations. %

\section{The model and the problem}

Let us consider a single quantum particle interacting with %
a potemtial field formed by very many mutually non-interacting %
plane-wave quantized excitations. The latter will be called %
``phonons'' while the particle ``electron'', although it will be %
spinless, or ``Brownian particle'' (BP), since its interaction %
with phonons may enforce it  to a random walk, at least in the %
thermodynamic limit and at non-zero temperature when the phonons %
can serve as thermostat with infinitely large total energy.

\subsection{Basic equations}

The Hamiltonian of the mentioned system will coincide with %
that of widely used simple models %
(see e.g. \cite{mah,fey,uw}) %
of quasi-free (conduction) electrons %
in movable (vibrating) crystal lattice:
\begin{eqnarray}
H\,=\,H_e+H_{ph} +H_{int}\,\,\,,\nonumber\\
H_e=\frac {p^2}{2m}\,\,\,, \,\,\,\,\,\, %
H_{ph}=\sum_k\, \hbar\omega_k\, a_k^\dagger a_k\,\,\,,\nonumber\\
H_{int}= \frac 1{\sqrt{\Omega}} \sum_k\, %
[\,c_k^* \,e^{\,ikr}a_k\,+\,c_k\,e^{-ikr} a_k^\dagger \,] %
\,\,\, \label{h}
\end{eqnarray}
Here\, $\,\Omega\,$ is the system's volume, %
so that wave vectors in the sums are separated %
by reciprocal volume $\,d^3k=(2\pi)^3/\Omega\,$\,, %
and
\begin{eqnarray}
c_{-k}\,=\,c_k^*\, \label{cc}
\end{eqnarray}
for any wave vector\, $\,k\,$.

Being interested first of all %
in statistics of  the electron's, or BP's, walk, %
we must consider the von Neumann evolution equation,
\begin{eqnarray}
i\hbar\, \dot{\rho} \,=\,[H,\rho]\,\equiv\, %
H\rho-\rho H %
\,\,\,, \label{e}
\end{eqnarray}
for full density matrix of the system,\, $\,\rho\,$. %
If at that we are interested first of all in finite %
temperatures, $\,T\equiv\beta^{-1}>0\,$, then it is %
reasonable to describe phonon dependence of the $\,\rho\,$ %
in terms of coherent states and the %
remarkable ``Sudarshan-Glauber %
P\,-representation'' \cite{sud,gl,glb} : %
\begin{eqnarray}
\rho\,=\, \int \cs\, P(\al,\al^*)\, d\al^*d\al %
\,\,\,, \label{pr}
\end{eqnarray}
where $\,\alpha =\{\alpha_k\}\,$, %
$\,\cs =\prod_k |\al_k\rangle\langle\al_k| \,$,\,  %
$\,d\al^*d\al=\prod_k d\al^*_kd\al_k = %
\prod_k d\Re\, \al _k\,d\Im\,\al _k\,$,\, and %
the ``quasi-probability density'' \cite{sud,gl,glb}, %
$\,P(\al,\al^*)\,$, in respect to the electron's variables %
still acts as an operator. %
Taking into account that (for any $\,k\,$)
\begin{eqnarray}
a\,\cs = \al\,\cs\,\,, \,\,\,\, %
a^\dagger\,\cs = (\pa +\al^*)\,\cs\,\,,\,\,\,\, \nonumber\\ %
\cs\,a = (\pa^*+\al)\,\cs\,\,,\,\,\,\, %
\cs\,a^\dagger = \al^* \,\cs\,\,, \label{r}
\end{eqnarray}
with
\[
\pa\equiv \frac {\pa}{\pa\al}\,\,\,,\,\,\,\,\, %
\pa^*\equiv \frac {\pa}{\pa\al^*}\,\,\,,
\]
one can transform Eq.\ref{e} into
\begin{eqnarray}
i\hbar\, \dot{P} = %
[H_e\,,\, P]\,+\, %
\sum_k\, \hbar\omega_k\,( \al^*_k \pa^*_k - %
\al_k \pa_k)\,P\,+\label{ep} \\
+\, %
\frac 1{\sqrt{\Omega}} \sum_k\,\{\, c_k^*\al_k %
 [e^{\,ikr},P] %
+ c_k\al_k^*\, [e^{-ikr},P] %
+ \,c_k^*\pa_k^* P\,e^{\,ikr} - %
c_k e^{-ikr}\pa_k P\,\}
\, \nonumber
\end{eqnarray}

Further, it will be convenient %
to treat an electron dependence of $\,P\,$ %
in the coordinate representation, %
introducing function\, $\,\langle r|\,P\,|r^{\,\prime}\rangle\,$\, %
instead of the operator $\,P\,$, with\, $\,|r\rangle\,$\, %
denoting eigenstates of the electron's coordinate operator \,$\,r\,$.\, %
Then it is convenient also to introduce %
new independent variables: %
\begin{eqnarray}
X=\frac {r+r^{\,\prime}}2\,\,, \,\,\,\, %
Y= r-r^{\,\prime}\,\,, \,\,\,\, \nonumber\\ %
A_k\,=\,\al_k\,e^{\,ikX}\,\,,\,\,\,\, %
A_k^*\,=\,\al_k^*\,e^{-ikX}\,\, \label{nv}
\end{eqnarray}
Designating the matrix element\, %
$\,\langle r|\,P\,|r^{\,\prime}\rangle\,$\, %
by the same symbol $\,P\,$,
\[
\langle r|\,P\,|r^{\,\prime}\rangle\, \equiv\, %
P(t,X,Y,A,A^*)\,\,\,,
\]
we obtain for it from Eq.\ref{ep} the equivalent equation %
\begin{eqnarray}
i\hbar\, \dot{P} = %
- \frac {\hbar^2}m \,\nabla_Y\, %
\left[\,\nabla_X\,+ \sum_k\,ik\,(A_k %
\pa_k-A_k^*\pa_k^*)\,\right]\,P \,+\,\nonumber\\ %
+\,\sum_k\, \hbar\omega_k\,( A^*_k \pa^*_k - %
A_k \pa_k)\,P\,+\label{epf} \\
+\, %
\frac 1{\sqrt{\Omega}} \sum_k\,[\,  %
(\,e^{\,ikY/2}-e^{-ikY/2})\, (\,c_k^*A_k- c_k A_k^*) %
+\,  e^{-ikY/2}\,(\,c_k^*\pa_k^* - c_k \pa_k\,)\,]\,P\, %
\,\, \nonumber
\end{eqnarray}
with
\[
\nabla_X=\frac {\pa}{\pa X}\,\,,\,\,\,\, %
\nabla_Y=\frac {\pa}{\pa Y}\,\,,\,\,\,\, %
\pa_k =\frac {\pa}{\pa A_k}\,\,,\,\,\,\, %
\pa_k^* =\frac {\pa}{\pa A_k^*}\,\, %
\]
The Fourier transformation\, %
$\,\int \exp{(-ipY/\hbar)}\, P\, d^3Y/(2\pi\hbar)^3\,$\, %
gives the Wigner representation with $\,P\,$ %
becoming literally ``quasi-probability density''.

\subsection{Initial conditions}

Being interested first of all in %
thermodynamically equilibrium electron's ``Brownian motion'', %
it is natural to start from such initial density matrix, - %
e.g. at  $\,t=0\,$, - %
which says that the phonon subsystem has canonical %
equilibrium probability %
distribution with given temperature $\,T=1/\beta\,$ %
while the electron is located in vicinity of some %
given point of space, - e.g. coordinate origin, - and %
has equilibrium Maxwellian momentum distribution %
corresponding to the same temperature. %
Such the density matrix can be expressed by formulas
\begin{eqnarray}
\rho(t=0)\,\propto\, |\Psi \rangle\langle\Psi  |\, %
\exp{(-\beta H_{ph})}\,\,\,,\label{ic}\\
\Psi (r)\,=\, \left(\frac {2mT}{\pi\hbar^2}\right)^{3/4}\, %
\exp{\left(-\frac %
{mT}{\hbar^2}\, r^2\right)}\,\,\, \nonumber
\end{eqnarray}
In order to rewrite this in the \,P-representation, %
notice \cite{glb} that for any particular phonon mode %
\begin{eqnarray}
(1-e^{-\beta\hbar\omega})\, \exp{(-\beta\hbar\omega\, %
a^\dagger a)}\,=\,\int \cs\, %
f(\omega,\al,\al^*)\, d\al^*d\al\,\,\,, \nonumber %
\end{eqnarray}
where
\begin{eqnarray}
f(\omega,\al,\al^*)\,\equiv\, \frac 1{\pi N(\omega)}\, %
\exp{\left[- \frac {|\al |^2}{N(\omega)}\right]}\,\,\,,\label{pe}\\
N(\omega)\,\equiv\, \frac %
1{\exp{(\beta\hbar\omega)}-1}\,\,\, \nonumber
\end{eqnarray}
Therefore initial condition (\ref{ic}) is equivalent to
\begin{eqnarray}
P(t=0)\,=\, \Psi (r)\Psi ^*(r^{\,\prime})\, %
\prod_k\, f(\omega_k,\al_k,\al_k^*)\,\,\,, \label{icp}
\end{eqnarray}
which is normalized distribution.

Obviously, such initial condition as (\ref{ic}) and (\ref{icp}) %
has serious defect:\, it neglects statistical correlations %
between electron and phonons inevitably arising from %
their interactions. However, this defect may be %
simultaneously its advantage:\, %
it helps to trace the development of the correlations %
during subsequent evolution directed by %
Eqs.\ref{e},\ref{ep} %
or \ref{epf}. Therefore, it is reasonable to accept %
such initial condition at least for a time.

\subsection{Marginal distributions}

The object of our direct interest is %
the marginal density matrix of electron,\, %
\begin{equation}
\rho_0\, =\, \prod_k\, \Tr_k\, \rho\,\,\,, %
\label{r0}
\end{equation}
- with\, $\,\Tr_k\,$ being trace over states of %
phonon mode\, $\,k\,$\,, - or corresponding marginal distribution %
\[
P_0\, =\,  \int P\, %
\prod_k\, d\al_k^*d\al_k\,  %
\,=  \int P\, \prod_k\, dA_k^*dA_k\,
\]
Alternative description of its evolution is presented %
by hierarchy of equations for the chain of various marginal, %
or particular, distributions
\[
P_n\, =\,  \int P\, %
\prod_{k\,\neq\,K_n}\, dA_k^*dA_k\, \,\,,
\]
where\, $\,K_n=\{k_1\dots k_n\}\,$\, is a set of %
wave vectors of $\,n\,$ {\bf different} phonon modes. %
Applying these integrations to Eqs.\ref{ep} or \ref{epf}, %
we come to hierarchy
\begin{eqnarray}
\dot{P}_n \, =\,  %
-\,\widehat{V} \nabla_X\, P_n\,+ %
\, \sum_{k\,\in\,K_n}\, [\, \widehat{L}_k\,+ %
\,\mu\,\widehat{A}_k %
\,+\,\mu\, \widehat{D}_k\, ]\, P_n\, + \label{epfn} \\
+\, \mu\, \sum_{q\,\neq\,K_n}\, \int \widehat{A}_q\, P_{n+1}  %
\, dA_q^*dA_q \,\,\,, \nonumber
\end{eqnarray}
where\, $\,P_{n+1}\,$\, corresponds to\, %
$\,K_{n+1}=\{K_n,q\}=\{k_1\dots k_n,q\}\,$\,, %
and we introduced operators %
\begin{eqnarray}
\widehat{V}\,\equiv\, -\frac {i\hbar}m\,\nabla_Y\,\,\,,\nonumber\\
\widehat{L}_k\,\equiv\, %
i\,(\omega_k\,-\,k\widehat{V})\,(A_k \pa_k - A^*_k \pa^*_k )\, %
\,\,, \label{ops}\\
\widehat{A}_k\,\equiv\, %
(i\hbar\sqrt{\Omega_0})^{-1}\,   %
(\,e^{\,ikY/2}-e^{-ikY/2})\, (\,c_k^*A_k- c_k A_k^*) %
\,\,\,, \nonumber\\ %
\widehat{D}_k\,\equiv\, (i\hbar\sqrt{\Omega_0})^{-1}\, %
e^{-ikY/2}\,(\,c_k^*\pa_k^* - c_k \pa_k\,)\, %
\,\,, \nonumber\\ %
\end{eqnarray}
and besides the small parameter\, %
\[
\mu\,\equiv \,\sqrt{\Omega_0/ %
\Omega}\,\,
\]
with %
$\,\Omega_0\,$ being some fixed ``microscopic'' volume.

It will be useful to treat these equations in terms of new %
functions $\,Q_n\,$ defined by
\[
P_n\,=\, Q_n\, \prod_{k\,\in\,K_n}\, f_k\,\,\,,
\]
with shortened notations
\[
\begin{array}{l}
f_k\,\equiv\, f(\omega_k,A_k,A_k^*)\, =\, %
(\pi N_k)^{-1}\, \exp{(-|A_k|^2/N_k)}\,\,\,,\\
N_k\,\equiv\,N(\omega_k)\,=\, %
1/[\,\exp{(\beta\hbar\omega_k)}-1\,]\,
\end{array}
\]
They undergo hierarchy of equations
\begin{eqnarray}
\dot{Q}_n \, =\,  %
-\,\widehat{V} \nabla_X\, Q_n\,+ %
\, \sum_{k\,\in\,K_n}\, [\, \widehat{L}_k\,+ %
\,\mu\,\widehat{A}_k %
\,+\,\mu\, f^{-1}_k\widehat{D}_k f_k\, ]\, Q_n\, %
+ \nonumber \\
+\, \mu\, \sum_{q\,\neq\,K_n}\, \int %
f_q\, \widehat{A}_q\, Q_{n+1}  %
\, dA_q^*dA_q \,= \label{eqn}\\
=\, -\,\widehat{V} \nabla_X\, Q_n\,+ %
\, \sum_{k\,\in\,K_n}\, [\, \widehat{L}_k\,+ %
\,\mu\,\widehat{B}_k %
\,+\,\mu\, \widehat{D}_k\, ]\, Q_n\, + \nonumber \\
+\, \mu\, \sum_{q\,\neq\,K_n}\, \int %
f_q\, \widehat{A}_q\, Q_{n+1}  %
\, dA_q^*dA_q \,\,\,, \nonumber
\end{eqnarray}
where
\begin{eqnarray}
\widehat{B}_k\,\equiv\, \widehat{A}_k\,+\, %
(i\hbar\sqrt{\Omega_0})^{-1}\, %
e^{-ikY/2}\,(\,c_k A_k^* - c_k^*A_k\,)/N_k\, %
 \label{b} %
\end{eqnarray}
Clearly, $\,Q_0\equiv P_0\,$, and initial %
conditions to Eqs.\ref{eqn} following from (\ref{icp}) are
\begin{eqnarray}
Q_n(t=0)\,=\, \Psi (r) \Psi ^*(r^{\,\prime})\,\, \label{icq}
\end{eqnarray}

Equations (\ref{epfn}) and (\ref{eqn}) are analogues of %
the BBGKY-like hierarchies of equations for BP in a classical %
ideal gas of atoms \cite{pro,ppro,lpro,tmf,ig} or generally %
in a simple fluid \cite{ppro,tmf,i1,p1,p3,i2}, %
with the last terms in (\ref{epfn}) and %
(\ref{eqn}) playing role %
of ``collision integrals''.

\,\,\,

The problem is again in investigation of properties of the BP's %
(electron's) distribution function $\,P_0=P_0(t,X,Y)\,$ %
what follow from Eqs.\ref{epfn} or \ref{eqn}.

\section{Thermodynamic limit and shortened hierarchy of %
evolution equations}

In order to grant to the BP (electron) large enough space %
for random walk and make the phonon subsystem to be %
as much as possible %
powerful thermostat, we must go to the %
thermodynamic limit\, $\,\Omega\rightarrow\infty\,$, %
\,$\,\mu\rightarrow 0\,$\,.

\subsection{Preliminary discussion}

Notice  that under this limit, %
firstly,  %
perturbation of any particular %
phonon mode by its interaction with electron is vanishingly %
small, so that
\begin{eqnarray}
P_n\,\rightarrow\,P_0\,\prod_{k\,\in\,K_n}\, %
f_k\,\,, \,\,\,\, %
Q_n\,\rightarrow\,Q_0=P_0\, \label{lim}
\end{eqnarray}
for all $\,n\,$. More precisely, %
the Eqs.\ref{epf},\ref{epfn} and \ref{eqn}, %
along with initial conditions (\ref{icp}) and (\ref{icq}), %
do imply that
\[
P\,=\, \left\{\sum_{s\,=\,0}^\infty\,\mu^s\, %
P^{(s)}\,\right\} \prod_k\, f_k\,\,\,,
\]
where\, $\,P^{(s)}\,$\, is a\, $\,s\,$-order polynomial %
of phonon variables\, $\,A_k^*,\,A_k\,$\,, with %
coefficients being some functions of $\,t,\,X\,$ %
and $\,Y\,$\,. %
Correspondingly,
\begin{eqnarray}
Q_n\,=\, \sum_{s\,=\,0}^\infty\,\mu^s\, %
P^{(s)}_n\,\,\,,\label{exp}
\end{eqnarray}
where\, $\,P^{(s)}_n\,$\, is a\, $\,s\,$-order polynomial %
of \, $\,A_k^*,\,A_k\,$\, with\, $\,k\,\in\,K_n\,$ and with %
coefficients representing some functions of $\,t,\,X,\,Y\,$ %
and $\,\mu\,$\,.

At the same time, secondly, %
because of finiteness of limits %
\[
\mu^2\sum_q\,\dots\, \rightarrow\,  %
\Omega_0 \int \dots\, d^3q/(2\pi)^3\,\,\,,
\]
- with dots replacing a ``good'' function  %
of $\,q\,$  %
(which is the case under natural restrictions on %
$\,|c_k|^2\,$), - %
summary perturbation of electron by all %
the phonon modes acquires at $\,\mu\rightarrow 0\,$ %
a constant value independent on $\,\mu\,$ and   %
indifferent to exclusion (or neglect) of any %
finite set of phonon modes. %
Therefore, coefficients of all %
the polynomials\, $\,P^{(s)}_n\,$\, %
in fact become independent on $\,\mu\,$.

These remarks show that, %
in agreement with (\ref{lim}), %
all $\,P_n^{(0)}\,$ with $\,n>0\,$  %
give no essential information in addition to %
one already contained in $\,P_0\,$. %
Actually, at $\,n=1\,$ a portion of %
new additional information is presented by  %
second term of expansion (\ref{exp}), i.e. by %
coefficients of the first-order polynomial %
$\,P_1^{(1)}\,$. Indeed, according to the %
first of Eqs.\ref{epfn} %
and \ref{eqn} (for $\,P_0\,$ and $\,Q_0\,$), %
at $\,\mu\rightarrow 0\,$ %
just $\,P_1^{(1)}\,$ directly and completely %
determines ``collision %
integrals'' in these equations %
(since $\,\mu^2\sum_q\dots \rightarrow %
\Omega_0 \int \dots d^3q/(2\pi)^3\,$ and, %
clearly, contributions from %
all $\,P_1^{(s)}\,$ with $\,s>1\,$ disappear %
in the thermodynamic limit).

Similarly, %
any of $\,P_n^{(1)}\,$ with $\,n>1\,$  %
gives no new information in comparison with %
that already contained in coefficients %
of $\,P_1^{(1)}\,$. %
Considering evolution of these coefficients, - %
with the help of second terms of %
(\ref{epfn}) or (\ref{eqn}) (at $\,n=1\,$), - %
one can see that at $\,\mu\rightarrow 0\,$ %
they are exactly determined, through corresponding %
components of collision integrals,  by coefficients of %
polynomial $\,P_2^{(2)}\,$. Or, to be more precise, %
four coefficient of bilinear part of $\,P_2^{(2)}\,$. %

Continuation of such reasonings leads to conclusion %
that hierarchy of functions, which unambiguously %
determine evolution of the electron's distribution %
function $\,P_0\,$, in the thermodynamic limit %
reduces to coefficients %
of {\bf purely multi-linear parts} of %
polynomials $\,P_n^{(n)}\,$ %
(which is easy understandable: since density %
of phonon modes in $\,k\,$-space tend to infinity, %
probability of strong excitation of one and the %
same mode becomes infinitely small as compared with %
probability of weak excitation of many different, %
let close in $\,k\,$-space, modes).

In order to confirm these statements, we have to %
extract from (\ref{epfn}) or (\ref{eqn}) a %
closed hierarchy of shortened equations for the mentioned %
coefficients. %

\subsection{Derivation of shortened equations}

First, let us define them formally:
\begin{eqnarray}
\Delta_n\,\equiv\, \lim_{\mu\,\rightarrow\,0}\,%
\frac 1{\mu^n}\left[\left(\, %
\prod_{k\,\in\,K_n}\, %
\pa_{k}^{\,\sigma_k}\,\right)\,  %
Q_n \right]_{A=A^*=0}\,\equiv\, \widehat{\Delta}_n\,Q_n %
\,\,\,, \label{d}
\end{eqnarray}
where\, $\,\sigma_k\,$\, can take one of two symbolic values, %
``\,+\,'' or ``\,-\,'', and %
we introduced new designations (more comfortable for further %
manipulations):
\[
\begin{array}{l}
A^-_k\,\equiv A_k\,\,\,, \,\,\,\,\, %
A^+_k\,\equiv A_k^*\,\,\,, \,\,\,\,\, %
\pa^-_k\,\equiv \pa_k\,\,\,, \,\,\,\,\, %
\pa^+_k\,\equiv \pa_k^*\,\,\,  %
\end{array}
\]
Applying such operation to $\,(n+1)$-th %
of Eqs.\ref{eqn} (for $\,Q_n\,$), consider different %
terms there separately.

For the second right-hand term one readily obtains
\begin{eqnarray}
\widehat{\Delta}_n \, \widehat{L}_k\,Q_n\, %
=\, -\,i \sigma_k\, %
(\omega_k\,-\,k\widehat{V})\, \Delta_n \,\,\,, \label{t2}
\end{eqnarray}
where $\,k\,$ is one of elements of the set $\,K_n\,$ %
from (\ref{d}). %

The third term needs in more careful consideration. %
Application of $\,\widehat{\Delta}_n\,$ to it gives
\begin{eqnarray}
\widehat{\Delta}_n \,\mu \widehat{B}_k\,Q_n\, %
=\,-\,(i\hbar\sqrt{\Omega_0})^{-1} \, %
\sigma_k\, c_k^{\sigma_k}\, \times \label{t3}\\ %
\times\, [\,e^{\,ikY/2}-e^{-ikY/2}(1+N^{-1}_k)\,]\, %
\widehat{\Delta}_{n-1}Q_n\,|_{A_k=A_k^*\,=0}\,\, %
\,\,, \nonumber
\end{eqnarray}
where
\[
\begin{array}{l}
c^-_k\,\equiv c_k^*\,\,\,, \,\,\,\,\, %
c^+_k\,\equiv c_k\,\,\,, \,
\end{array}
\]
and, clearly,\, $\,\widehat{\Delta}_{n-1}\,$\, %
corresponds to the set\, %
$\,K_{n-1}=K_n\ominus k\,$\,. %

Next, to express $\,\widehat{\Delta}_{n-1}Q_n\,$ %
in (\ref{t3}) through functions (\ref{d}), %
we should take into account, firstly, %
that by the $\,Q_n\,$-s definition %
\begin{eqnarray}
\int Q_n\, f_q\, \,dA_q^*dA_q\, =\, Q_{n-1}\, %
\,\,\,\,\,\,\,\,\,\, (K_n=K_{n-1}\oplus q)\, %
\, \label{qq}
\end{eqnarray}
Therefore
\begin{eqnarray}
\int \left[\,\frac 1{\mu^{n-1}} %
\prod_{k\,\in\,K_{n-1}} %
\pa_{k}^{\,\sigma_k} \,\, Q_n\,\right]\, f_q\, %
\,dA_q^*dA_q\, =\, %
\frac 1{\mu^{n-1}} \prod_{k\,\in\,K_{n-1}} %
\pa_{k}^{\,\sigma_k}\,\, Q_{n-1}\, %
\, \label{dqq1}
\end{eqnarray}
Secondly, turning here all\, $\,A_k^*,\,A_k\,$ with %
$\,k\,\in\,K_{n-1}\,$ into zeros,  %
in view of the expansion (\ref{exp}) for the square %
bracket on the left  %
we can write
\begin{eqnarray}
[\,\dots\,]_{A_{K_{n-1}}=A_{K_{n-1}}^*\,=0}\, =\, %
[\,\dots\,]_{A=A^*\,=0}\, %
+\,\sum_{s\,=\,1}^\infty\,\mu^s\,p^{(s)}(A_q,A_q^*) %
\,\,\,,  \label{dqq2}
\end{eqnarray}
where\, $\,A_{K_{n-1}}\,$\, means any of $\,A_k\,$ %
with $\,k\in K_{n-1}\,$, %
while $\,A\,$ any of all phonon variables %
with wave vectors from $\,K_n\,$, and\, %
$\,p^{\,(s)}(A_q,A_q^*)\,$\, is $\,s\,$-order %
homogeneous polynomial resulting from %
$\,P_n^{\,(n-1+s)}\,$.

Combining (\ref{dqq1}) and %
(\ref{dqq2}), in the limit $\,\mu\rightarrow 0\,$ %
we come to equality
\begin{eqnarray}
\widehat{\Delta}_{n-1}Q_n\,|_{A_q=A_q^*\,=0}\, %
=\,\Delta_{n-1} \,\,\, \,\,\,\,\, %
(K_n=K_{n-1}\oplus q)\,  \label{dqq}
\end{eqnarray}
With its help Eq.\ref{t3} yields
\begin{eqnarray}
\widehat{\Delta}_n \,\mu \widehat{B}_k\,Q_n\, %
=-\,(i\hbar\sqrt{\Omega_0})^{-1}  %
\sigma_k\, c_k^{\sigma_k}\, %
[\,e^{\,ikY/2}-e^{-ikY/2}(1+N^{-1}_k)]\, %
\Delta_{n-1}\,\,\, %
\, \label{t31}
\end{eqnarray}
(with\, $\,K_{n-1}=K_n\ominus k\,$).

Now, consider action of $\,\widehat{\Delta}_n \,$ %
onto fourth right-hand term of (\ref{eqn}): %
\begin{eqnarray}
\widehat{\Delta}_n \, \mu\,\widehat{D}_k\,Q_n\,=\, %
\lim_{\mu\,\rightarrow,0}\,\mu^2 \, %
(i\hbar\sqrt{\Omega_0})^{-1}\, %
e^{-ikY/2}\,\times \nonumber\\
\times\, \left[ \frac 1{\mu^{n+1}} %
\,(\,c_k^*\pa_k^* - c_k \pa_k\,) %
\prod_{q\,\in\,K_{n}} %
\pa_{q}^{\,\sigma_q} \,
Q_n\,\right]_{A=A^*=0} %
\,\, \nonumber %
\end{eqnarray}
Evidently, the square bracket here has a finite limit %
at $\,\mu\rightarrow 0\,$, again thanks to (\ref{exp}). %
Therefore the total expression tends to zero, so that %
\begin{eqnarray}
\widehat{\Delta}_n \, \mu\,\widehat{D}_k\,Q_n\,=\,0\, %
\,\label{t4}
\end{eqnarray}

At last, consider action of $\,\widehat{\Delta}_n \,$ %
onto ``collision integrals in (\ref{eqn}). %
At that, we should use identities
\begin{eqnarray}
\int f_q\,A_q\,[\,\dots\,]\, dA_q^*dA_q\,=\, %
\int \{-N_q\,\pa_q^*\,f_q\,\}\,A_q\, %
[\,\dots\,]\, dA_q^*dA_q\, %
=\, \nonumber\\
=\,N_q \int f_q\,\pa_q^*\,[\,\dots\,]\, %
dA_q^*dA_q\,\,\,,\nonumber\\ %
\int f_q\,A_q^*\,[\,\dots\,]\, dA_q^*dA_q\,=\, %
N_q \int f_q\,\pa_q\,[\,\dots\,]\, dA_q^*dA_q\,\,\,,\label{fn} %
\end{eqnarray}
where square bracket means arbitrary function %
of the phonon variables. %
Hence, collision integrals can be rewritten as follows, %
\begin{eqnarray}
\widehat{C}_{n+1}\,Q_{n+1}\,\equiv\, %
\mu\, \sum_{q\,\neq\,K_n}\, \int %
f_q\, \widehat{A}_q\, Q_{n+1}  %
\, dA_q^*dA_q \,=\,\nonumber\\ %
=\,\mu\, (i\hbar\sqrt{\Omega_0})^{-1}\, %
\sum_{q\,\neq\,K_n}\,(\,e^{\,iqY/2}-e^{-iqY/2})\, %
N_q\, \times \label{ci1}\\
\times\, \int %
f_q\, (\,c_q^*\pa_q^*- c_q \pa_q)\, Q_{n+1}  %
\, dA_q^*dA_q \,\,\,, \nonumber
\end{eqnarray}
and
\begin{eqnarray}
\widehat{\Delta}_n\,\widehat{C}_{n+1}\,Q_{n+1}\,=\,\nonumber\\ %
=\,(i\hbar\sqrt{\Omega_0})^{-1}\, %
\lim_{\mu\,\rightarrow\,0}\, \mu^2\,  %
\sum_{q\,\neq\,K_n}\,(\,e^{\,iqY/2}-e^{-iqY/2})\, %
N_q\, \times\nonumber\\
\times\, \int f_q\, %
\sum_{\sigma_q \in \{+,-\}}\, %
\sigma_q\,c_q^{-\sigma_q}\, %
\left[ \frac 1{\mu^{n+1}}\,\pa_q^{\,\sigma_q} %
\prod_{k\,\in\,K_n}\, \pa_k^{\,\sigma_k}\, %
Q_{n+1}\,\right]_{A_{K_n}=A^*_{K_n}\,=0}  %
 dA_q^*dA_q \,\, \nonumber %
\end{eqnarray}
Here $\,A_{K_n}\,$ means any of $\,A_k\,$ with $\,k\in K_n\,$. %
According to (\ref{exp}), the square bracket here %
behaves at $\,\mu\,\rightarrow\,0\,$ in full analogy %
with that in (\ref{dqq2}),
\begin{eqnarray}
[\,\dots\,]_{A_{K_n}=A^*_{K_n}\,=0}\, =\, %
[\,\dots\,]_{A=A^*\,=0}\, %
+\,\sum_{s\,=\,1}^\infty\,\mu^s\, %
\widetilde{p}^{\,(s)}(A_q,A_q^*) %
\,\,\,,  \nonumber
\end{eqnarray}
with polynomials\, %
$\,\widetilde{p}^{\,(s)}(A_q,A_q^*)\,$\, %
now produced from $\,P_{n+1}^{\,(n+1+s)}\,$. %

Consequently, after performing the limit  %
and transforming the sum over $\,q\,$ into %
integral, we obtain
\begin{eqnarray}
\widehat{\Delta}_n\,\widehat{C}_{n+1}\,Q_{n+1}\,=\,\nonumber\\ %
=\,\frac {\sqrt{\Omega_0}}{i\hbar}\,  %
\sum_{\sigma \in \{+,-\}}\, \sigma %
\int c_q^{-\sigma}\, %
(\,e^{\,iqY/2}-e^{-iqY/2})\, N_q \,\,  %
\Delta_{n+1}\, \,\frac {d^3q}{(2\pi)^3}\, %
\,\,, \label{t5}
\end{eqnarray}
where\, $\,K_{n+1}=K_n\oplus q\,$\, %
and \,$\,\sigma\,$\, corresponds to \,$\,q\,$ %
(so that one can replace %
dummy variable $\,q\,$ by $\,k_{n+1}\,$ %
and its pair  $\,\sigma\,$ by $\,\sigma_{n+1}\,$). %

As the result of relations (\ref{t2}), (\ref{t31}), %
(\ref{t4}) and (\ref{t5}), we come  %
under the thermodynamic limit to the following exact %
short-cut (nevertheless, infinite) hierarchy of %
evolution equations:
\begin{eqnarray}
\dot{\Delta}_n \, =\,  %
-\,\widehat{V} \nabla_X\, \Delta_n\, %
-\,i\sum_{k\,\in\,K_n}\,  \sigma_k\, %
(\omega_k\,-\,k\widehat{V})\, \Delta_n %
\,-\,\nonumber\\ %
-\,(i\hbar\sqrt{\Omega_0})^{-1} %
\sum_{k\,\in\,K_n}\, %
\sigma_k\, c_k^{\sigma_k}\, %
[\,e^{\,ikY/2}-e^{-ikY/2}(1+N^{-1}_k)\,]\, %
\Delta_{n-1}\,+ \label{edn}\\ %
+ \,\frac {\sqrt{\Omega_0}}{i\hbar}\,  %
\sum_{\sigma\in \{+,-\}}\, \sigma %
\int c_q^{-\sigma}\, %
(\,e^{\,iqY/2}-e^{-iqY/2})\, N_q \,\,  %
\Delta_{n+1}\, \,\frac {d^3q}{(2\pi)^3}\, %
\,\, \nonumber
\end{eqnarray}
Here $\,K_{n-1}=K_n\ominus k\,$ and %
$\,\Sigma_{\,n-1}=\Sigma_{\,n}\ominus \sigma_k\,$ %
in the third term on the right and %
$\,K_{n+1}=K_n\oplus q\,$ and %
$\,\Sigma_{\,n+1}=\Sigma_{\,n}\oplus \sigma\,$ %
in the last tern %
(collision integral), with $\,\Sigma_{\,n}\,$ %
denoting sets of the two-fold indices paired with %
wave vectors.

Recall that all  wave vectors in any of the sets $\,K_n\,$ %
have appeared different one from another, so that  %
``collision integral'' in the last term, %
strictly speaking, does not include (infinitely small %
neighborhoods of) $\,n\,$ points $\,q\in K_n\,$. %
However, both %
the $\,\Delta_n\,$'s definition (\ref{d}) %
and structure of Eqs.\ref{edn} allow to extend %
functions $\,\Delta_n\,$ to coinciding wave %
vectors by continuity.

Initial conditions to Eqs.\ref{edn}, what literally %
follow from  (\ref{ic}) or (\ref{icq}),
\begin{eqnarray}
\Delta_n(t=0)\,=\, \delta_{n\,0}\, %
\Psi (r)\Psi ^*(r^{\,\prime})\,\,\,, \label{icd}
\end{eqnarray}
show that there is no need %
in any a priori assumptions about these functions.

As for the electron's probability distribution function %
$\,\Delta_0=P_0\,$, in essence, %
Eqs.\ref{edn} altogether form an exact time-non-local %
kinetic equation for it. At that, its part of the %
initial conditions can be chosen arbitrarily.

\subsection{Stationary solution, %
equilibrium distribution, %
and thermodynamically %
improved initial conditions}

The Eqs.\ref{edn} by their  %
derivation describe an electron %
in infinitely large phonon (or generally boson) %
thermostat formed by %
continuum of modes each being %
initially in canonical %
equilibrium state. %
Therefore, firstly, %
stationary solution of Eqs.\ref{edn}, - %
to be denoted as $\,\Delta^{eq}_n\,$, - %
must represent thermodynamically equilibrium distribution %
of the electron's momentum %
under fully uncertain electron's %
position, so that
\[
\nabla_X\, \Delta^{eq}_n\,=\,\dot{\Delta}^{eq}_n\,=\,0\, %
\]
Such solution can not be strictly normalized to unit, -  %
in respect to both momentum and %
coordinate of the electron, - %
but it can be normalized in respect to momentum only:
\[
\Delta^{eq}_0(Y=0)\,=\,1\,\,\,
\]
Secondly, time-dependent (strictly normalized) %
solution of Eqs.\ref{edn} %
must tend to such stationary equilibrium solution %
in the sense that
\[
\lim_{t\,\rightarrow\,\infty}\, %
\int \Delta_n\, d^3X\,=\, %
\Delta^{eq}_n\,\,\,
\]

The functions $\,\Delta^{eq}_n\,$ %
contain complete information about %
equilibrium statistical correlations %
between electron's momentum and thermostat. %
Hence, if we want to take into account these %
inevitable correlations from the very beginning, %
we should replace initial conditions (\ref{icd}) %
by
\begin{eqnarray}
\Delta_n(t=0)\,=\, W_0(X)\, \Delta^{eq}_n\, %
\,\,, \label{icd1}
\end{eqnarray}
where a normalized probability %
density $\,W_0(X)\,$ introduces initial %
localization of the electron in %
configurational space. %
Then time-dependent solution of Eqs.\ref{edn} %
will highlight specific %
``historical'' statistical correlations %
\cite{ppro,tmf} %
arising between current state of the thermostat and %
summary path of BP (electron) %
during all the time interval $\,(0,t)\,$. %
It may be said that the ``equilibrium  %
correlations'' describe momentary electron-phonons %
interaction while the ``historical'' ones %
describe a heritage of past interactions.

\section{Equivalent representations and %
generating functional of electron-phonon correlations}

A suitable change of the variables $\,\Delta_n\,$ %
at $\,n>0\,$ helps to remove the formally arbitrary %
parameter $\,\Omega_0\,$ and fashion Eqs.\ref{edn} %
to a more visual form. For instance, introducing, %
instead of $\,\Delta\,$, functions $\,D_n\,$ %
defined by
\begin{eqnarray}
\Delta_n\,=\,D_n\, %
\prod_{k\,\in\,K_n}\, s_{k\,\sigma_k}\, %
\,\,, \,\,\,\,\, %
\, s_{k\,\sigma}\,=\, %
-\frac {c_k^{\,\sigma} %
}{\hbar\omega_k\,N_k \sqrt{\Omega_0}} %
\, \,\, \label{d2}
\end{eqnarray}
at $\,n>0\,$ and $\,D_0=\Delta_0\,$, %
we transform Eqs.\ref{edn} into
\begin{eqnarray}
\dot{D}_n \, =\,  %
-\,\widehat{V} \nabla_X\, D_n\, %
-\,i\sum_{k\,\in\,K_n}\,  \sigma_k\, %
(\omega_k\,-\,k\widehat{V})\, D_n %
\,-\,\nonumber\\ %
-\,i \sum_{k\,\in\,K_n}\, %
\sigma_k\,\omega_k \, %
[\,e^{\,ikY/2}N_k\,-\,e^{-ikY/2}(N_k+1)\,]\, %
D_{n-1}\,+ \label{edn1}\\ %
+ \,i \sum_{\sigma\in \{+,-\}}\, \sigma %
\int \frac {|c_q|^2}{\hbar^{\,2}\omega_k}\, %
\,(\,e^{\,iqY/2}-e^{-iqY/2})\, \,  %
D_{n+1}\, \,\frac {d^3q}{(2\pi)^3}\, %
\,\,, \nonumber
\end{eqnarray}
with the same meaning of $\,\sigma\,$ and %
$\,q\,$ as before.

\subsection{Kinetic equation and infinite %
hierarchy of irreducible correlations}

Physical and statistical %
interpretation of just made change becomes clear %
if notice, firstly, %
that, - as one can prove %
in the spirit of above consideration, - %
at different wave vectors %
definition (\ref{d}) %
is equivalent to  %
\begin{eqnarray}
\Delta_n\,=\, \prod_{k\,\in\,K_n}\, %
\frac 1{N_k}\, %
\lim_{\mu\,\rightarrow\,0}\,%
\frac 1{\mu^n} \int P_n\, \prod_{k\,\in\,K_n}\, %
A_{k}^{-\sigma_k}\,\, %
dA^*_kdA_k\,\,\, \label{d1}
\end{eqnarray}
Secodly, if the electron was immovable %
(``infinitely hard'') then perturbation %
of all the phonon amplitudes would reduce %
to their constant shifts by
\begin{eqnarray}
\overline{A}^\sigma_k\,=\,- %
\frac {c_k^\sigma }{\hbar\omega_k\,\sqrt{\Omega}}\, %
\,=\,\mu s_{k\,\sigma}N_k\, %
\nonumber
\end{eqnarray}
Comparison of this expression with (\ref{d2}) %
shows that
\begin{eqnarray}
D_n\,=\, \int  %
P_n\, \prod_{k\,\in\,K_n}\, %
\frac {A_{k}^{-\sigma_k}} %
{\overline{A}_{k}^{-\sigma_k}}\, %
\,dA^*_kdA_k\,\,\,, \nonumber
\end{eqnarray}
that is $\,D_n\,$ ($\,n>0\,$) represent %
joint perturbation of $\,n\,$ (different) %
phonon modes measured in natural units %
of their separate static perturbations. %
Under the Wigner representation of %
$\,D_n\,$'s electron dependence %
this relation can be rewritten as %
\begin{eqnarray}
D_n(t,X,p,K_n,\Sigma_n)\,=\, D_0(t,X,p)\,  %
\left\langle \,\prod_{k\,\in\,K_n}\, %
\frac {A_{k}^{-\sigma_k}(t)} %
{\overline{A}_{k}^{-\sigma_k}} %
\right\rangle_{X,p}\,\,\,, \label{ds}
\end{eqnarray}
where the angle brackets mean conditional %
average of (quasi-classic) phonon amplitudes %
at given electron's position and %
momentum. %

At that, our basic definition (\ref{d}) %
prompts that functions $\,D_n\,$ ($\,n>0\,$) %
indeed must be extended to %
(zero-measure subspaces of) %
coinciding  wave vectors by continuity.

From the Hamiltonian (\ref{h}) it is %
seen that, in case of immovable (``static'') %
electron %
all phonon modes would be statistically independent %
one on another, %
and all the statistical moments in (\ref{ds}) %
factored  into products of first-order moments %
(average values). %
In reality, however, electron's motion couples different %
phonon modes one with another making them statistically %
dependent. Therefore their statistical moments consist of %
not only average values but also second-\, and %
higher-order irreducible correlations %
(cumulants). They %
reflect both current and old interactions %
(figuratively speaking, ``Correlations = Interactions %
+ History''). %
The Eqs.\ref{edn1} exactly describe %
influence of ``current'' onto %
dynamics and ``old'' onto statistics %
of electron motion (for detail explanations %
of nature of ``historical'' correlations %
see e.g. \cite{i1,bk12,bk3,i2,last,kr}) . %

\subsection{Generating functional}

It may be convenient to accumulate all the %
components of Eqs.\ref{edn1} into a single object, %
such as vector\, $\,\{D_0\,,\,D_1\,,\,\dots\,\}\,$ %
in the corresponding Fock space or the generating %
functional:
\begin{eqnarray}
\mathcal{D}\,\equiv\, D_0\,+\sum_{n\,=\,1} %
^\infty \,\frac 1{n!} %
\sum_{\sigma_1\dots \sigma_n}\,\int %
D_n\, \prod_{j\,=\,1}^n\, z_{\sigma_j}(k_j)\,d^3k_j\, %
\,\, \nonumber
\end{eqnarray}
so that
\[
D_n\,=\, \left[\,\prod_{j\,=\,1}^n\,\frac %
{\delta}{\delta z_{\sigma_j}(k_j)}\,\,\, \mathcal{D}\, %
\right]_{z\,=\,0}
\]
In terms of this functional the whole %
infinite hierarchy (\ref{edn1}) reduces to %
single equation  %
\begin{eqnarray}
\dot{\mathcal{D}} \, =\,  %
-\,\widehat{V} \nabla_X\, \mathcal{D}\, %
+\,\sum_\sigma \int d^3k\,\,\, z_\sigma(k)\, %
\widehat{L}_{k\,\sigma}\,\frac %
\delta{\delta z_\sigma(k)}\,\,\mathcal{D}\, %
+\, \label{fe}\\
+\, \sum_\sigma \int d^3k\,\,\, z_\sigma(k)\, %
\widehat{B}_{k\,\sigma}\, %
\,\mathcal{D}\,+\, %
\sum_\sigma \int d^3k\,\,\, %
\widehat{A}_{k\,\sigma}\,\frac %
\delta{\delta z_\sigma(k)}\,\,\mathcal{D}\, %
\,\equiv\, \nonumber\\
\equiv\, %
-\,\widehat{V} \nabla_X\, \mathcal{D}\, %
+\, \widehat{\mathcal{L}} %
\left\{z,\frac \delta{\delta z} %
\right\}\,\mathcal{D}\, %
\,\,, \nonumber
\end{eqnarray}
where
\begin{eqnarray}
\widehat{L}_{k\,\sigma}\,\equiv\, %
-\,i  \sigma\, %
(\omega_k\,-\,k\widehat{V})\,\,\,, \nonumber\\
\widehat{B}_{k\,\sigma}\,\equiv\, %
-\,i \sigma\,\omega_k \, %
[\,e^{\,ikY/2}N_k\,-\,e^{-ikY/2}(N_k+1)\,]\, %
\,\,, \label{fops} \\
\widehat{A}_{k\,\sigma}\,\equiv\, %
i \sigma %
\frac {|c_k|^2}{\hbar^{\,2}\omega_k\, (2\pi)^3}\, %
\,(\,e^{\,ikY/2}-e^{-ikY/2})\,  %
\,\,, \nonumber
\end{eqnarray}
and $\,\widehat{V}\,$ is electron's %
velocity operator defined in (\ref{ops}).

Quite similarly we can introduce generating %
functional $\,\Delta\,$ for the %
correlation functions $\,\Delta_n\,$ and %
then write out similar evolution equation for it. %
Clearly, both they are connected with $\,\mathcal{D}\,$ %
and Eq.\ref{fe} through simple scale transformation %
of the test function $\,z=z_\sigma(k)\,$,
\begin{eqnarray}
\Delta\{z\}\,=\, \mathcal{D}\{sz\}\,\,\,, \nonumber\\
\dot{\Delta}\,=\, %
-\,\widehat{V} \nabla_X\, \Delta\, %
+\, \widehat{\mathcal{L}} %
\left\{sz,\frac \delta{s\,\delta z} %
\right\}\,\Delta\, %
\,\,, \label{fe0}
\end{eqnarray}
with above  %
introduced multiplier $\,s=s_{k\,\sigma}\,$.

The functional evolution %
equation (\ref{fe}) is direct formal %
analogue of equations for BP in fluids %
investigated in \cite{pro,ppro,lpro,tmf,ig}. %
Investigation of Eq.\ref{fe} or Eq.\ref{fe0} %
may be subject of future works. %
At the rest of this one we confine ourselves %
by discussion of some consequences and   %
modifications of the Eqs.\ref{fe} and \ref{fe0}. %

\section{Generalizations and special cases}

\subsection{Other particles and external potentials}

In presence of additional ``external'' particles, %
interacting with the phonon thermostat %
and distributed with given   %
density $\,\nu(r)\,$, the Hamiltonian (\ref{h}) %
transforms as
\begin{eqnarray}
H\,\Rightarrow \,H\,+\,H_a\,\,\,, \,\,\,\,\,\, %
H_a\,=\,
\frac 1{\sqrt{\Omega}} \sum_k\, %
[\,c_k^* \,\nu_k^* a_k\,+\,c_k\,\nu_k a_k^\dagger \,] %
\,\,\,, \label{ha}
\end{eqnarray}
where
\[
\nu_k\,\equiv\, \int e^{-ikr} \nu(r)\, d^3r\,\,\,
\]
Then after repeating, %
for new Hamiltonian (\ref{ha}), %
the  above expounded derivation of %
shortened equations,  %
one easy comes to Eq.\ref{fe} with %
modified third right-side term: %
\begin{eqnarray}
\widehat{\mathcal{L}}\,\,\Rightarrow\,\, %
\widehat{\mathcal{L}}^{\,\prime}\,\equiv\,
\widehat{\mathcal{L}}\,+\, %
\sum_\sigma \int d^3k\,\,\, z_\sigma(k)\, %
[\,i \sigma\,\omega_k \, %
\nu_k^{\,\sigma}\,\exp{(i\sigma kX)}\,]\, %
\,\,, \label{fops1}
\end{eqnarray}
with\, $\,\nu_k^+\equiv \nu_k\,$\,,\, %
$\,\nu_k^-\equiv \nu_k^*\,$\,. %
This addition to evolution operator %
$\,\widehat{\mathcal{L}}\,$ %
does not include the electron momentum %
related operators $\,\widehat{V}\,$ %
and $\,Y\,$. Therefore it does not create %
additional electron-phonon %
correlations (instead producing %
momentum-independent shifts of mean values %
$\,\langle A_k^\sigma \rangle\,$ only). %
Hence, it can be removed by %
representing solution\, $\,\mathcal{D}^\prime\,$\, %
to the modified evolution equation, %
\begin{eqnarray}
\dot{\mathcal{D}}^\prime\,\,=\, %
-\,\widehat{V} \nabla_X\, \mathcal{D}^\prime\, %
+\, \widehat{\mathcal{L}}^\prime %
\left\{z,\frac \delta{\delta z} %
\right\}\,\mathcal{D}^\prime\,\,\,, %
\label{fem}
\end{eqnarray}
in the form
\begin{eqnarray}
\mathcal{D}^\prime\,=\, \mathcal{D}\,\, %
\exp{\left[\,\sum_\sigma\int %
z_\sigma(k)\,\nu_k^\sigma \, %
e^{\,i\sigma\,kX}\,\, %
d^3k\,\right]\, \label{cv} }
\end{eqnarray}
Then  %
a simple algebra %
turns the Eq.\ref{fem} into %
\begin{eqnarray}
\dot{\mathcal{D}}\,\,=\, %
-\,\widehat{V} \nabla_X\,\mathcal{D}\, +\, %
\frac {U(r)-U(r^{\,\prime})}{i\hbar}\,\, %
\mathcal{D}\, +\, %
\widehat{\mathcal{L}} %
\left\{z,\frac \delta{\delta z} %
\right\}\,\mathcal{D}\,\,\,, %
\label{feu}
\end{eqnarray}
where\, $\,r\,,\,r^{\,\prime}\,= %
\,X\,\pm\,Y/2\,$\, and %
\begin{eqnarray}
U(r)\,=\, -\int \frac {|c_k|^2}{\hbar\omega_k}\, %
[\,\nu_k\,e^{\,ikr}\,+\, \nu_k^*\,e^{-ikr}\,]\, %
\frac {d^3k}{(2\pi)^3}\,\, \label{u}
\end{eqnarray}
is potential of effective (mediated by phonons) %
interaction of our electron with the  ``external'' %
particles.

This result is not significant in itself, %
for it follows already directly from the %
Hamiltonian (\ref{ha}), with the help %
of boson operator amplitudes shifting %
\[
A_k \,+\,\frac {c_k\nu_k}{\hbar %
\omega_k\sqrt{\Omega}}\,\Rightarrow\,A_k\,\,\,,\,\,\,\,\,\, %
A_k^* \,+\,\frac {c_k^*\nu^*_k}{\hbar %
\omega_k\sqrt{\Omega}}\,\Rightarrow\,A_k^*\,\,\,
\]
Nevertheless, it is important for us since it %
verifies faultlessness of our derivation of %
Eqs.\ref{edn} or Eqs.\ref{edn1} and thus Eqs.\ref{fe}.

\subsection{Many phonon branches %
(boson thermostats)}

One of generalizations of the model (\ref{h}), %
which may be necessary in its applications, %
is accounting for presence of several %
types, or ``branches'', of %
phonon modes, with different dispersion %
laws and/or couplings to the electron. %
Obviously, to perform such a generalization, %
it is sufficient to impart an additional %
index, - let ``b'', - to the phonon amplitudes %
$\,A_k,A_k^*\,$, frequencies %
$\,\omega_k\,$ and coupling constants %
$\,c_k^\sigma\,$, the sets %
of $\,n\,$ such indices to the correlation functions %
$\,\Delta_n\,$ and $\,D_n\,$ and,  %
correspondingly, the same additional index %
to the test function $\,z_\sigma(k)\,$ and %
all operators (\ref{fops}), %
so that
\[
\sum_\sigma \int d^3k\,\,\dots\,\,\Rightarrow %
\,\, \sum_b %
\sum_\sigma \int d^3k\,\,\dots\,\,
\]
in Eqs.\ref{fe} and other functional %
evolution equations.

Equivalently, we can divide %
the corresponding total functional %
evolution operator, %
$\,\widehat{\mathcal{L}}\,$, into sum %
of operators %
$\,\widehat{\mathcal{L}}^{(b)}\,$ %
each relating to one of the %
phonon branches, so that the Eq.\ref{fe} %
takes form %
\begin{eqnarray}
\dot{\mathcal{D}} \, =\,  %
-\,\widehat{V} \nabla_X\, \mathcal{D}\, %
+\,\sum_b\,
\widehat{\mathcal{L}}^{\,(b)} %
\left\{z_b,\frac \delta{\delta z_b} %
\right\}\,\mathcal{D}\, %
\,\, \label{fes}
\end{eqnarray}
Advantage of this representation %
is in that it naturally extends %
to the case of different %
temperatures of various phonon branches %
(in other words, boson thermostats), %
which is obviously allowed by our above %
analysis. %

\subsection{Application to electron in %
static disorder}

Let us consider a quantum particle (``electron'') %
in Gaussian static random potential $\,\Phi(r)\,$ with %
zero mean value and %
correlation function
\begin{eqnarray}
\langle\, \Phi(r_1)\,\Phi(r_2)\,\rangle\,=\, %
\int |c_k|^2\,e^{\,ik(r_1-r_2)}\, d^3k/(2\pi)^3\, %
\label{cf}
\end{eqnarray}
We assume that %
the electron's density matrix, $\,\rho_e\,$, %
at some time moment, $\,t=0\,$, is known, %
$\,\rho_e(t=0)=\rho_{e0}\,$, and want to know %
its mean value at $\,t\neq 0\,$,
\begin{eqnarray}
\overline{\rho}_e(t)\,=\, %
\langle\, e^{-iHt/\hbar}\, %
\rho_{e0}\, e^{\,iHt/\hbar}\, \rangle\,\, %
\,, \label{mdm}
\end{eqnarray}
where\, $\, H\,=\,p^2/2m\,+\,\Phi(r)\,$\, %
is Hamiltonian of the system.

In calculations of this quantity %
the so-called operator representation %
of random processes and fields may be %
useful \cite{or}. For instance, any %
statistical moment of Gaussian field with %
auto-correlation (\ref{cf}) can be represented %
as
\begin{eqnarray}
\langle\, \prod_j\,\Phi(r_j)\,\rangle\,=\, %
\langle 0|\,\prod_j\,\widehat{\Phi}(r_j) %
\,|0\rangle\,\,\,, \nonumber
\end{eqnarray}
where
\begin{eqnarray}
\widehat{\Phi}(r)\,=\,
\frac 1{\sqrt{\Omega}} \sum_k\, %
[\,c_k^* \,e^{\,ikr}a_k\,+\,c_k\,e^{-ikr} a_k^\dagger \,] %
\,\,\, \nonumber
\end{eqnarray}
is quantum boson field operator with coefficients %
satisfying (\ref{cc}), $\,\Omega\rightarrow\infty\,$, %
and\, $\,|0\rangle\,$\, is the ``vacuum state''.
Therefore instead of (\ref{mdm}) one can write
\begin{eqnarray}
\overline{\rho}_e(t)\,=\, %
\langle 0|\, e^{-iHt/\hbar}\, %
\rho_{e0}\, e^{\,iHt/\hbar}\, |0\rangle\, %
\,\,, \nonumber
\end{eqnarray}
now with\, %
$\,H\,=\,p^2/2m\,+\,\widehat{\Phi}(r)\,$\,. %

From here it follows that
\begin{eqnarray}
\langle \Psi|\,\overline{\rho}_e(-t)\, %
|\Psi\rangle\,=\, %
\Tr_e\,\rho_{e0}\,\{\,\Tr_b\, %
e^{-iHt/\hbar}\, |0\rangle |\Psi\rangle %
\langle\Psi| \langle 0|\, %
e^{\,iHt/\hbar}\,\}\, %
\,\,, \label{tr}
\end{eqnarray}
where\, $\,|\Psi\rangle\,$ is arbitrary %
electron state and\, $\,\Tr_e\,,\,\Tr_b\,$ %
mean traces over electron and boson degrees %
of freedom, respectively. %
Obviously, the expression in the braces %
is particular and peculiar case of %
the expression (\ref{r0}) arising from it %
under limits\, $\,T\rightarrow 0\,$\, %
and\, $\,\omega_k\rightarrow 0\,$\,. %
At that, the first of these limits %
should be performed before the second, %
since transition to vacuum initial %
state of boson subsystem presumes %
that $\,N_k\rightarrow 0\,$. %

To realize such ordered limits in Eqs.\ref{edn}, %
we firstly have to rescale functions $\,\Delta_n\,$ %
by
\begin{eqnarray}
\Delta_n\,=\,D_n\, %
\prod_{k\,\in\,K_n}\, s_{k\,\sigma_k}\, %
\,\,, \,\,\,\,\, %
\, s_{k\,\sigma}\,=\, %
-\frac {c_k^{\,\sigma} %
}{\hbar\,N_k \sqrt{\Omega_0}} %
\, \,\,, \nonumber
\end{eqnarray}
instead of (\ref{d2}), then %
set $\,N_k=0\,$ and after that %
$\,\omega_k=0\,$. This results %
in equations
\begin{eqnarray}
\dot{D}_n (t,X,Y,K,\Sigma)\, =\,  %
[-\nabla_X\, %
+\,i\sum_{k\,\in\,K}\,  \sigma_k\, %
k\,]\,\widehat{V}\, %
D_n(t,X,Y,K,\Sigma)\,+ \nonumber\\ %
+\,i \sum_{k\,\in\,K}\, %
\sigma_k\, e^{-ikY/2}\, %
D_{n-1}(t,X,Y,K\ominus k, %
\Sigma\ominus \sigma_k)\,+ \label{edn2}\\ %
+ \,i \sum_{\sigma}\, \sigma %
\int \frac {|c_q|^2}{\hbar^{\,2}}\, %
\,(\,e^{\,iqY/2}-e^{-iqY/2})\, \,  %
D_{n+1}(t,X,Y,K\oplus q, %
\Sigma\oplus\sigma)\, \,\frac {d^3q}{(2\pi)^3}\, %
\,\, \nonumber
\end{eqnarray}
(with designations slightly different %
from that in Eqs.\ref{edn1} but clearly %
equivalent to them). %
According to (\ref{mdm}) and (\ref{tr}), %
if $\,D_n(t)\,$ is solution of these equations %
under initial conditions %
\begin{eqnarray}
D_n(t=0)\,=\,\delta_{n\,0}\, %
|\Psi\rangle\langle\Psi|\,\,\,,\label{icd2}
\end{eqnarray}
then
\begin{eqnarray}
\langle \Psi|\,\overline{\rho}_e(-t)\, %
|\Psi\rangle\,=\, %
\Tr_e\,\rho_{e0}\,D_0(t)\, %
\,\, \label{tr1}
\end{eqnarray}
In particular, at %
$\,\rho_{e0}=|\Psi_0\rangle\langle\Psi_0|\,$ %
we obtain expression, %
\begin{eqnarray}
\langle\, |\langle\Psi| %
e^{\,iHt/\hbar}|\Psi_0\rangle|^2\, %
\rangle\, %
=\, %
\langle \Psi_0| \,D_0(t)\, %
|\Psi_0\rangle\,
\,\,, \label{tr2}
\end{eqnarray}
for mean probability of electron's transition %
between two given states $\,\Psi_0\,$ and $\,\Psi\,$. %

Because of the ``Anderson %
localization'' of a part of electron states %
in random potential \cite{an,bs,ik}, %
the quantity (\ref{tr2}) has non-zero limit %
at $\,t\rightarrow\pm\infty\,$ %
depending on both %
$\,\Psi_0\,$ and $\,\Psi\,$. %
This fact means that stationary solution %
of Eqs.\ref{edn2} is not unique but strongly %
degenerated %
(consequence of that now the ``thermostat'' has zero %
internal energy and thus no ability %
to ``thermalize'' electron). %
These multiple stationary solutions can %
be used to represent such (``eigen'') %
electron states which are correlated with %
the disorder (in contrast to a priory %
given states $\,\Psi_0\,$ and $\,\Psi\,$). %

\subsection{Naive ``weak-coupling limit'' %
and conventional kinetic equation}

Returning to the case when\, $\,T\neq 0\,$ and %
$\,\omega_k\neq 0\,$, let us assume that some %
sufficient smallness of the %
coupling constants $\,|c_k|^2\,$,  %
e.g. in the sense of
\[
\gamma\,\equiv\, %
\int \frac {|c_k|^2}{\hbar^2\omega_k}\, %
\frac {d^3k}{(2\pi)^3}\,\ll\, \frac T\hbar \,\,\,,
\]
gives us rights to truncate the %
hierarchy (\ref{edn}) or, equivalently, %
(\ref{edn1}) at its second level by setting %
$\,\Delta_2=0\,$ or $\,D_2=0\,$, %
i.e. neglecting third-order correlations (between %
electron and two phonon modes) and higher-order ones. %
Then, after transition to the Wigner representation, %
so that %
\[
\exp{(\pm ikY/2)}\,F(Y)\,\Rightarrow\, %
F(p \mp \hbar k/2)\,\,\,, %
\]
we come to equation
\begin{eqnarray}
\dot{D_0}(t,X,p)\,=\, -V\nabla_X\,D_0(t,X,p)\, %
+\, \sum_\sigma \int %
\frac {d^3k}{(2\pi)^3}\, \frac {|c_k|^2}{\hbar^2} %
\int_0^t d\tau\, \times \,\nonumber\\
\times\, %
\{ %
\exp{[ -i\sigma ( %
\omega_k- kV +\frac {\hbar k^2}{2m} %
)\,\tau\,]}\,\times\nonumber\\ %
\times\, %
[N_k\,D_0(t-\tau,X-\tau %
(V-\frac {\hbar k}{2m}),p-\hbar k) %
- \nonumber\\ %
-\,(N_k+1)\, D_0(t-\tau,X-\tau %
(V-\frac {\hbar k}{2m}),p)] %
\,+\,\nonumber\\ %
+\,\exp{[ -i\sigma ( %
\omega_k- kV -\frac {\hbar k^2}{2m} %
)\,\tau\,]}\, %
\times\nonumber\\ %
\times\, %
[(N_k+1)\,D_0(t-\tau,X-\tau %
(V+ \frac {\hbar k}{2m}),p+\hbar k) %
\,-\, \nonumber\\
-\, N_k\, D_0(t-\tau,X-\tau %
(V+\frac {\hbar k}{2m}),p)] %
\} %
\,\,\,,\label{ke0}
\end{eqnarray}
where\, $\,V=p/m\,$. %
Further standard assumptions about %
spatial dependence (i.e. $\,X\,$-dependence) %
of the density matrix and standard reasonings %
(or ``conjurations''), %
along with formal identity %
\[
\lim_{t\,\rightarrow\,\infty}\, %
\sum_\sigma %
\int_0^t \exp{[\,\sigma\,(\,\dots\,)\,\tau\,]}\, %
d\tau\,=\,2\pi\, \delta(\,\dots\,)\,\,\,,
\]
help to transform (\ref{ke0}) into what is called %
``kinetic equation'',  %
\begin{eqnarray}
\dot{D_0}(t,X,p)\,=\, -V\nabla_X\,D_0(t,X,p)\, %
+\,\label{ke}\\ %
+ \,\frac {2\pi}{\hbar^2}
 \int %
|c_k|^2\, %
\{\, %
\delta(\omega_k- kV +\frac {\hbar k^2}{2m}) %
\,\times \nonumber\\ %
\times \, %
[\,N_k\,D_0(t,X,p-\hbar k) %
- \,(N_k+1)\, D_0(t,X,p)]\,+\,\nonumber\\ %
\,+\,\, %
\delta(\omega_k- kV -\frac {\hbar k^2}{2m}) %
\,\times\nonumber\\
\times\, [(N_k+1)\,D_0(t,X,p+\hbar k) %
\,-\, N_k\, D_0(t,X,p)\,] %
\,\} %
\, \frac {d^3k}{(2\pi)^3}\, %
\,\,\,, \nonumber
\end{eqnarray}
which is one-electron version of more general %
kinetic equation for electrons in vibrating %
crystal lattice \cite{lp}.

Thus we have demonstrated reducibility of our %
exact evolution equations to standard %
kinetic models.

Unfortunately, ``to break is not to make''! %
The truncation of hierarchy of electron-many-phonon %
correlations is much easier than revealing of  %
their actual predestination %
and physical sense. %
Therefore let us go to %
summing up %
our actual results.

\section{Discussion and resume}

The kinetic equation (\ref{ke}) %
takes into account %
only simplest mutually non-correlated %
(mutually incoherent) %
electron transitions, each involving %
one phonon. %
In reality, however, electron's interaction %
with any particular %
phonon mode $\,k\,$, - described by %
operators\, %
$\,a_k\exp{[ikr(t)]}\,$ and %
$\,a_k^\dagger \exp{[-ikr(t)]}\,$\, %
in the Heisenberg %
representation over other modes, - %
depends on random electron's %
path\, $\,r(t)\,$\, %
determined by simultaneous interactions %
with all these modes. %
Hence, there are also various %
multi-stage electron's transitions which %
involve mutually correlated %
(mutually coherent) acts of absorption %
or irradiation of two, three or more phonons. %
Hierarchy of the above introduced %
correlation functions $\,\Delta_n\,$ %
and $\,D_n\,$ %
just describes %
how these %
many-particle processes %
```dress'' the electron and constitute %
statistics %
of its random walk.

Analogous correlated %
many-particle process were considered %
in \cite{kmg}, where quantum amplitude of any %
particular electron transition was influenced by %
electric voltage fluctuations due to %
simultaneously realizing transitions %
of other electrons. %
There it was demonstrated that %
the correlated transitions %
are responsible for low-frequency %
(``quasi-static'', or 1/f-type) fluctuations %
of one-electron ``transition probabilities'' %
and consequently electric conductance. %

Here, we performed %
similar but rigorous analysis %
of electron (quantum particle) interacting %
with infinitely many phonon modes (boson thermostat) %
and showed  %
possibility of shortened but complete %
description of this system %
in terms of the mentioned %
electron-many-phonon correlation functions %
whose arguments include electron variables %
and phonon indices only. %

Main formal result of the present paper is %
derivation of exact hierarchy %
of evolution equations %
for these correlation function and thus %
for electron's density matrix. %
That are Eqs.\ref{edn} or Eqs.\ref{edn1} %
and their generating functional %
equivalent, Eq.\ref{fe}. %
For the best of my knowledge, %
such equations never earlier were in use. %

It should be underlined that our recipe of %
exact shortened description can be extended to %
(generally relativistic) ``dressed'' electron %
in electromagnetic field %
and to many-electron (or, generally, %
many-fermion) systems.

The correlated many-particle processes %
(or ``kinetic events'', %
or ``dynamic clusters'', etc.) %
in classical statistical mechanics %
of fluids %
\cite{pro,ppro,lpro,tmf,ig,p1,p3} %
and dielectric crystals \cite{i3} %
are under our attention %
after the work \cite{i1} %
where for the first time it was shown %
that a fluid particle has no definite %
``probability of collisions'', %
diffusivity and mobility. %
In other words, any of such %
characteristics of particle's random motion 
undergoes scaleless 1/f-type low-frequency %
fluctuations. %
The theory discovers them when it does not neglect %
$\,n\,$-particle statistical correlations %
which can arise between particles participating %
in chains (clusters) of $\,n-1\,$ %
dynamically connected consecutive %
collisions. %
Infiniteness of  variety %
of such chains means originality of any %
phase trajectory of the %
system \cite{tmf,kr} and %
impossibility of its fully adequate %
description in terms %
of two-particle events only, %
even in case of arbitrarily dilute %
gas \cite{pro,lpro,tmf,ig}.

Close ideas concerning random motion of %
electrons (or holes) in electric conductors, - %
e.g.  intrinsic or weakly doped semiconductors,  - %
were formulated already in \cite{bk12,bk3,pr1,pr2} %
(for later formulations see also \cite{i2,last}), %
with the aim to suggest principal explanation %
of electric 1/f-noise observed in various conductors %
\cite{bk3} and usually manifesting itself as %
1/f fluctuations in mobilities of charge carriers %
\cite{bk3,hkv,dyre,tv}. %
The present work is first step from the %
phenomenological theory of electronic %
1/f-noise \cite{bk12,bk3,pr1,pr2} to its %
microscopic statistical-mechanical theory.

In this respect, I would like to emphasize %
close conformity of the above obtained %
Eqs.\ref{edn1} and \ref{fe} for ``Brownian particle'' %
(electron) in ideal phonon gas and previously %
obtained  %
evolution equations for many-particle correlations %
``dressing'' a molecular Brownian particle in %
usual classical ideal gas \cite{pro,lpro,tmf,ig}. %
Solution of %
Eqs.\ref{edn1} or \ref{fe} can be represented by %
an (operator-valued and branching) %
infinite continued fraction %
similar to that in \cite{pro}. %
This conformity prompts that the Eqs.\ref{edn1} %
or, equivalently, \ref{fe} also hide 1/f fluctuations %
of diffusivity and mobility. %
Their visualization must %
be second step to the desired theory.

At that, of course, none truncation of Eqs.\ref{edn1} %
is appropriate. The generating-functional formulation, %
(\ref{fe}), of Eqs.\ref{edn1} make it obvious that all %
floors of the corresponding continued fraction, - i.e. %
all the electron-$\,n\,$-phonon irreducible correlations, - %
have equal rights and significance, since all %
they are represented in Eq.\ref{fe} by only three %
simple terms. And it is so at %
arbitrarily weak electron-phonon coupling. %
Weaker coupling implies only %
that upper frequency bound of %
1/f fluctuations of rate of electron relaxation %
becomes lower proportionally to the rate itself, %
and no more. %
Hence, one needs in adequate methods of operating %
with the whole electron's correlation ``coat''. %
This will be third step to valuable theory %
of electron-phonon interaction.

\,\,\,

I would like to acknowledge Prof. Yu.\,V.\,Medvedev %
for useful discussions.

%

\,\,\,

---------------------------

\,\,\,

\end{document}